# It depends: Varieties of defining growth dependence

Anja Janischewski [a *], Katharina Bohnenberger [b,c], Matthias Kranke [d], Tobias Vogel [e], Riwan Driouich [f], Tobias Froese [g], Stefanie Gerold [h], Raphael Kaufmann [i], Lorenz Keyßer [j], Jannis Niethammer [k], Christopher Olk [l], Matthias Schmelzer [m,n], Aslı Yürük [o], Steffen Lange [p]

[a] *Faculty of Economics and Business Administration, Chemnitz University of Technology, Germany*
[b] *German Institute for Interdisciplinary Social Policy Research, University of Bremen, SOCIUM, Germany*
[c] *Institute for Socio-Economics, University of Duisburg-Essen, Germany*
[d] *Freiburg Institute for Advanced Studies (FRIAS), University of Freiburg, Germany*
[e] *Department for Philosophy, Politics and Economics, Faculty of Economy and Society, Witten/Herdecke University, Germany*
[f] *Institut de Ciència i Tecnologia Ambientals (ICTA), Universitat Autònoma de Barcelona (UAB), Spain*
[g] *Chair for Corporate Sustainability, ESCP Business School, Germany*
[h] *Institute of Philosophy and Social Science, Brandenburg University of Technology Cottbus-Senftenberg, Germany*
[i] *ZOE Institute for Future-Fit Economies, Germany*
[j] *Institute of Geography and Sustainability, Faculty of Geosciences and Environment, University of Lausanne, Switzerland*
[k] *ICLEI European Secretariat, Germany*
[l] *Otto Suhr Institute for Political Science, Freie Universität Berlin, Germany*
[m] *Norbert-Elias-Center for Transformation Design and Research, University of Flensburg, Germany*
[n] *Friedrich-Schiller-University Jena, Germany*
[o] *Urban Transformation and Global Change Laboratory (TURBA), Universitat Oberta Catalunya, Spain*
[p] *Centre for Pluralist Economics, University of Siegen, Germany*

* Corresponding author at: *Faculty of Economics and Business Administration, Chemnitz University of Technology, Germany*
*E-mail address*: contact@anja-janischewski.com (Anja Janischewski).

**Abstract:**

Many socio-economic systems require positive economic growth rates to function properly. Given uncertainty about future growth rates and increasing evidence that economic growth is a driver of social and environmental crises, these growth dependencies pose serious societal challenges. In recent years, more and more researchers have thus tried to identify growth-dependent systems and develop policies to reduce their growth dependence. However, the concept of "growth dependence" still lacks a consistent definition and operationalization, which impedes more systematic empirical and theoretical research. This article proposes a simple but powerful framework for defining and operationalizing the concept of "growth dependence" across socio-economic systems. We provide a general definition consisting of four components that can be specified for different empirical cases: (1) the system under investigation, (2) the unit of measurement of growth, (3) the level of growth and (4) the relevant functions or properties of the system under investigation. According to our general definition, a socio-economic system is growth-dependent if it requires a long-term positive growth rate in terms of a unit of economic measurement to maintain all its functions or properties that are relevant within the chosen normative framework. To illustrate the usefulness of our scheme, we apply it to three areas at the heart of the existing literature on growth dependence: employment, social insurance systems and public finance. These case studies demonstrate that whether or not a system is growth-dependent hinges not only on the empirical properties of the system itself but also on the specification of the concept of growth dependence. Our framework enables coherent, robust and effective definitions and research questions, fostering comparability of findings across different cases and disciplines. Better research can lead to better policies for reducing growth dependence and thus achieving stable and sustainable economies.

# 1. Introduction

Modern societies have embraced and institutionalized the operating principle that unless one goes forward, one goes backward. In fact, many argue that contemporary societies would not function without economic growth (Malmaeus & Alfredsson 2017; Strunz & Schindler 2018). Seen in this light, growth (in)dependence is the elephant in the room when it comes to the polarized debate on green growth versus post-growth. Yet while those promoting green growth contend that the economy must expand to avoid catastrophic social impacts, many post-growth advocates downplay or overlook the potential socio-economic repercussions of their own proposals. A better understanding of the causes and dynamics of growth dependence, as well as possibilities for instituting less growth-dependent societal systems, is thus critical for sustainability transitions (see Corlet Walker et al. 2021, 2024). However, understanding the causes of growth dependence to implement transitional policies presupposes the systematic identification of growth dependent systems, which, in turn, requires a coherent operationalization of the concept.

In this article, we propose a novel framework for operationalizing the concept of *growth dependence* and, by implication, also that of *growth independence*. A careful operationalization allows one to clearly distinguish whether a specific system is growth-dependent or not, without a priori knowledge about the causal mechanisms of the phenomenon. Furthermore, we illustrate that the assessment of growth dependence may depend on the definition and operationalization used. Our endeavor is motivated by two main considerations regarding the role of economic growth in modern societies. First, a long term trend of low or zero growth – a phenomenon known as "secular stagnation" – constitutes a risk to societal welfare, especially in already affluent economies, which tend to have more modest growth rates (Teulings and Baldwin 2014; Jackson 2018; European Environment Agency 2021, pp. 39–40). Second, recent evidence lends little support to the hope for near-term absolute decoupling of economic growth from environmental pressures in line with internationally agreed climate targets and broader parameters of ecosystem health (Haberl et al. 2020; Hickel and Kallis 2020; Parrique et al. 2019). As long as the strive for continuous economic growth is institutionally locked in, societies tread on a dangerous path toward eroding the ecological foundations of the life support systems of humans and non-humans alike – or they confront serious social, economic and political crises in the wake of implementing environmental policies that effectively limit growth. For both these reasons, greater engagement with the dynamics and forms of growth dependence is overdue, especially if one follows a precautionary principle of post-growth planning (Petschow et al. 2020).

If low or no growth is socio-economically disastrous but continued growth ecologically devastating, reducing growth dependence offers a pragmatic way forward across a wide range of otherwise deadlocked policy debates. Discussions of growth dependence do not necessarily problematize economic growth as such, but the reliance of certain socio-economic systems on growth to meet human needs (Büchs 2021; Corlet Walker et al. 2024). Thus, both academic and policy-oriented contributions have recently advanced proposals that aim at sustainable well-being and societal resilience independent of economic growth in an age of planetary polycrisis (European Environment Agency 2021; Jackson 2016; Seidl & Zahrnt 2010; Petschow et al. 2020; Schmelzer et al. 2022). Growth independence can also be seen as a precondition for realizing a state of "a-growth" (van den Bergh 2017) and has been discussed in approaches, including "doughnut economics" (Raworth 2017), the "economy for the common good" (Felber 2015), "well-being economies" (Trebeck & Williams 2019), and specific progressive versions of a Green New Deal (Aronoff et al. 2019; Pettifor 2019).

The article proceeds as follows. In section 2, we review the relevant literature on growth dependence and related concepts. In section 3, we present our novel analytical framework for operationalizing growth dependence within a scientifically sound approach that comprises four components, namely (1) the system under investigation, (2) the measuring unit, (3) the level of growth and (4) relevant functions and properties. We illustrate the framework's value-added in section 4 by using employment, social insurance systems and public finance as examples of how growth dependence is institutionalized in multiple societal sectors. We discuss interconnections of the three examples and outline avenues for future research in section 5 before concluding in section 6.

## 2. Literature and context

Since the publication of *The Limits to Growth* report in 1972, ecological economists, among others, have underlined path dependencies in the pursuit of continued economic growth that obstruct transformations towards sustainable economies (Douthwaite 1993; Jackson 2016). Particularly the science-activist field of degrowth has criticized the "addiction to growth" (Costanza 2022), the "growth paradigm" (Schmelzer 2015) or "growthism" (Daly 2019) as an encompassing ideology that shapes the mental, socio-cultural and material structures of most contemporary societies. Accordingly, the last decade has seen various attempts at developing more nuanced scientific assessments of these path dependencies and their underlying growth mechanisms in order to grasp them theoretically, analyze them empirically, and, potentially, overcome them politically. Among relevant conceptualizations that have emerged from these undertakings are "dynamic stabilization" (Rosa 2016), "growth imperatives" (Richters & Siemoneit 2019), "growth drivers" and "growth dependencies" (Jackson 2016). In this article, we focus on "growth dependence" as a comprehensive concept that can be further specified and integrate most of these related concepts without equating them in terms of definition.

So far, the associated debates have remained scattered, incomplete or restricted to certain fields of application. Research on growth dependencies is sparse although it can be found in fields as diverse as business administration (Gebauer 2018), social innovation research (Tschumi et al. 2021), and urban planning (Rydin 2013). The academic discussion is arguably most advanced in the field of (sustainable) welfare systems, where Corlet Walker et al. (2024) and Wiman (2024) have made crucial contributions to conceptualizing, operationalizing and analyzing growth dependence (see also Büchs 2021; Koch 2022). Similarly, Oberholzer (2023) models macroeconomic stability under a post-growth scenario, yet without referring to growth dependence as a concept.

In contrast to economic growth theory as an established research field that inquires the factors that cause economic growth (see Bassanini & Scarpetta 2001), scholarship on the social stabilization functions of growth is nascent. Notably, Corlet Walker et al. (2024) propose five dimensions of growth dependence that cover the relevant aspects of the phenomenon occurring in the literature. In contrast, we do not focus on systemizing the aspects of existing phenomena, but on providing a minimal set of dimensions that a researcher has to specify to operationalize the concept, a-priori to open-ended research. So far, the literature lacks a conceptual framework that is specific enough to enable such open-ended empirical assessment of growth dependence (specific conditions) while being broad enough to cover all relevant instances of growth dependence and avoid definitional exclusions (general scope). For such an integrative, comparative discussion, clear definitions are required. As Ostrom (1986, p. 4) puts it: "No scientific field can advance far if the participants do not share a common understanding of key terms in their field." We build on this reminder to develop an analytical understanding of "growth dependence" by proposing an overarching concept that captures the stabilizing functions of economic growth. In a first step, therefore, we map growth dependence onto the related and partially overlapping concepts of "growth drivers", "growth imperatives", and "dynamic stabilization". In a second step, we review the range of existing approaches to growth dependence and highlight their ambiguities to sharpen the concept internally.

*Growth drivers* can be understood as factors that tend to stimulate economic growth, but at the same time do not penalize its absence (Strunz & Schindler 2018; see also Corlet Walker et al. 2024). For example, the well-established economic model of growth considers labor-productive technological progress through innovation as a fundamental growth driver (Solow 1956; Romer 1990). Here, the concept seeks to explain how positive growth trends occur (Petschow et al. 2020, ch.3). The term is also frequented by the comparative economics literature, where it is understood as unique elements that, while not part of the total income, affect the expansion of its individual components (Kohler and Stockhammer, 2021; Baccaro and Hadziabdic, 2022). Its meaning is closer to "growth dependence" in the sense of penalizing the absence of growth, when denoting a strong disposition of economic agents toward growth and a social pressure to make growth happen (Richters & Siemoneit 2019). Conversely, the perception – or narrative construction – that our society is dependent on growth can lead to decisions that strengthen growth drivers (see Gumbert et al. 2022).

*Growth imperatives* generally refer to the danger of extremely detrimental consequences in the absence of growth (Kettner & Vogel 2021). Richters and Siemoneit (2019, p. 129, original emphasis) explicitly highlight systemic pressures "to avoid *existential consequences*". More specifically, the existence of growth imperatives means that a zero-growth state is not feasible because falling below a certain level of positive growth rates would usher in a contraction process with negative growth rates (Binswanger 2006, Binswanger 2019). Similarly, various concepts of "growth dependence" emphasize the existential consequences of insufficient growth for the maintenance of social systems (Tschumi 2021, p.119; Mayer et al. 2021, p. 219; Schmelzer et al. 2022, p.207). Some also identify significant negative consequences or harm (Corlet Walker et al. 2021; Bohnenberger & Fritz 2020) through which "certain core aspects of human wellbeing become compromised" (Corlet Walker et al. 2024, p. 1). Others, however, define the scope of growth dependencies more broadly, simply stating that without growth, beneficial functions or widely accepted social objectives cannot be realized to the same extent (Wagner & Lange 2021; Petschow et al. 2020). In contrast to growth imperatives, growth dependence can also exist when lacking or too little growth undermines a certain standard of social functioning, even if it does not trigger hardship for individuals or threaten the persistence of social systems. A further difference is that the concept of "growth dependence" emphasizes the adverse consequences of lacking or too little growth without presupposing a reversal to negative growth rates, as explained above.

*Dynamic stabilization*, popularized by sociologist Hartmut Rosa and colleagues (Rosa 2016; Rosa et al. 2017), combines growth drivers and growth imperatives within a theory of growth-dependent societies. The term specifically refers to how modern societies reproduce themselves through processes of economic growth, innovation intensification and acceleration, without which they destabilize (Rosa 2016, p. 673). Dynamic stabilization intertwines the growth drivers underlying the cultural-material promises of an accelerated individual life with the growth imperatives of a capitalist economy (Rosa 2016, pp. 679–680, 684–687). Technical innovations help to cope with life's increasing pace while at the same time driving growth and accelerating socio-structural change. The latter in turn makes living conditions more dynamic and fast-paced, triggering a cycle of dynamic stabilization (Rosa 2012, p. 309).

In contrast to these approaches, the definitions of *growth dependence* present in the literature consist of three main aspects. First, many accounts suggest that the absence of economic growth threatens the maintenance of societal functions (see e.g. Lange 2018; Petschow et al. 2020; Tschumi et al. 2021, p. 119; Schmelzer et al. 2022, p. 168; Corlet Walker et al. 2024). Some forms of growth dependence may constitute existential threats, but the maintenance of societal functions is the lowest common denominator and a necessary condition for growth dependence. Researchers should thus specify on a case-by-case basis which function is affected, in which way, and why this effect is problematic.

Second, a large part of the literature underlines the need for *continued* growth because otherwise negative consequences will materialize (Schmid 2021 p.65; Schmelzer et al. 2022, p. 168; Bohnenberger & Fritz 2020; Lange 2022). In this respect, a necessary component of growth dependence is the requirement of not only a temporary growth spurt but an ongoing process of economic expansion. In an application to pension funds, Wiman (2024) focuses on relatively "worse economic outcomes in a no-growth scenario compared to a growth scenario" (Wiman 2024, p.2). How growth is measured, however, varies across approaches. While most conceptualizations refer to growth of the gross domestic product (GDP), others foreground dependence on microeconomic measures, such as the size of a company in general or firm's revenues (Gebauer 2018, p. 240; Meyer et al. 2021, p. 227).[1] Furthermore, the modifier *continuous* in "continued growth" leaves room for interpretation regarding the required growth rates or the allowed degree of fluctuations along a growth trend.

Third and finally, definitions refer to a social entity to which growth dependence is attributed. However, the conceptualization and scope of the analyzed entities differs widely, including "society" as a whole (Tschumi et al. 2021, p. 119), "areas of society" (Petschow et al. 2020, p. 91), "institutions" (Lange 2022; Petschow et al. 2020, p. 91), or "infrastructures" (Schmelzer et al. 2022, p. 149–50), and combinations thereof. A term that captures this broad spectrum is thus required. For example, "areas of society", "institutions" or "infrastructures" exclude society as a whole and are too narrow to capture complex logics, such as Rosa's "dynamic stabilization". On the other end of the spectrum, the term "society" itself cannot denote areas of society such as labor markets or social security systems. The concept of "institutions" is ambiguous, being used in the sense of "organizations", "rules" or "social norms" (see e.g. Ostrom 1986; Hodgson 2006), so that the term "growth-dependent institutions" can describe various entities. Accurate research needs to clearly specify and delimit the social entity considered to be growth-dependent. In the next section, we thus propose using *systems* as a concept that is both fairly encompassing and specific enough for operationalization across contexts.

In summary, the literature refers to growth dependence as a state in which a social entity requires some form of growth to prevent undesired consequences. Beyond this broad understanding, conceptualizations remain mostly vague. A notable exception is Corlet Walker et al. (2024), who outline five general dimensions of growth dependence. Our framework partly builds upon these dimensions. Overall, the literature exhibits inconsistencies along multiple dimensions, as discussed above. In particular, the system under investigation, the measuring unit of growth, the meaning of growth and the consequences of insufficient growth are often either not defined or defined in different ways. Consequently, existing applications of the concept frequently capture only a fraction of the different manifestations of growth dependence. We present a non-exhaustive list of existing definitions in Appendix A.

We provide a framework for operationalization that can be applied across disciplines and is agnostic to the underlying mechanisms. Defining a concept means striking a certain balance between elasticity and rigor: A useful concept is one that can be specified with a certain degree of flexibility to account for a variety of empirical cases and contexts. At the same time, the definition should be stable and systematic enough to ensure comparability across cases and contexts. The presented framework strikes this balance by providing a broad, structured, but also bounded space of growth dependence specifications. We loosely draw on the method of concept analysis, focusing on identifying the attributes and boundaries that define a concept (Walker & Avant 2019). The framework does not presuppose the type of the underlying causal mechanism of growth dependence which can refer to structural, institutionalized systems as well as to cultures and ideologies as belief systems or to individual preferences and mentalities as psychological systems, even though our examples only refer to structural, institutionalized systems.[2] Furthermore, the framework serves as a template for building specific definitions for concrete research questions by spelling out the defining characteristics of growth dependence, rather than providing a list of observed phenomena. It shall help researchers to construct "intensional definitions", which list the necessary and sufficient conditions for something to fall under the defined concept (Kockaert & Steurs 2015, p. 11). These definitions of growth dependence can then be the starting point for open-ended analyses of the growth dependence of concrete socio-economic systems.

---

[1] Some authors identify the term growth with increasing energy and material throughput in the context of current environmental problems (see e.g. Hickel 2021). Our approach, however, focuses on economic measuring units, not least to emphasize the difference between growth dependence and the disputed ability to decouple economic growth from its environmental impacts.

# 3. Definition and framework

In existing literature, growth dependence generally means that some socio-economic system requires growth of an economic variable to avoid potentially detrimental consequences for the system. Consequently, to define and operationalize this concept, researchers have to at least specify the following four aspects: (1) the system under investigation, (2) the economic variable, (3) the meaning of *growth* and (4) the *consequences* for the system. More concretely, we understand *consequences* as the non-maintenance of functions or properties of the system that are relevant for human well-being. Our general definition of growth dependence is as follows:

> A socio-economic system is growth-dependent if it requires a positive long-term growth rate of a unit of economic measurement to maintain all its relevant functions or properties**.**

In other words, a socio-economic system is growth-dependent[3] if at least one system function or one relevant property cannot be maintained when growth rates[4] are too low – regardless of whether they are still positive, zero or even negative. What function or property counts as *relevant* in this definition has to be consciously evaluated by researchers, based on normative criteria for human well-being that should be made explicit. *Growth independence*, in turn, is given if the system under investigation does not require positive growth rates in the specified measuring unit to sustain its relevant functions and properties. The four main elements of the general definition must be specified to yield concrete, case-specific definitions of growth dependence (see Figure 1 for a schematic overview).

---

[2] The links between growth dependence and cultures, ideologies, and mentalities are too multifaceted to be adequately addressed in this essay. For example, cultures, ideologies, and mentalities may be expressions of existing growth-dependent systems, or they may generate growth-dependencies themselves, e.g., for a political system, through the voting behavior of the people. Furthermore, they may be a determinant of how less growth translates into negative consequences, or they may simply represent unfounded claims about negative consequences.

[3] We acknowledge the multiple meanings of the term "dependence". A similar but more general meaning of an item A depending on an item B is that B affects A, i.e. changes in B result in changes in A (Oxford Learner's Dictionaries). This meaning is relevant for gradual effects such as the negative correlation between growth and employment rates in Okun's law. However, in line with existing definitions, we focus on the meaning of "dependence" to be "a state of needing something or someone, esp. in order to continue existing or operating" (Cambridge Dictionary).

References: https://www.oxfordlearnersdictionaries.com/definition/english/dependence**,** https://dictionary.cambridge.org/dictionary/english/dependency.

[4] While growth often refers to macroeconomic measuring units only, the framework presented here is also applicable to the dependence on growth of firm- or sector-level measuring units, such as revenue or production input.

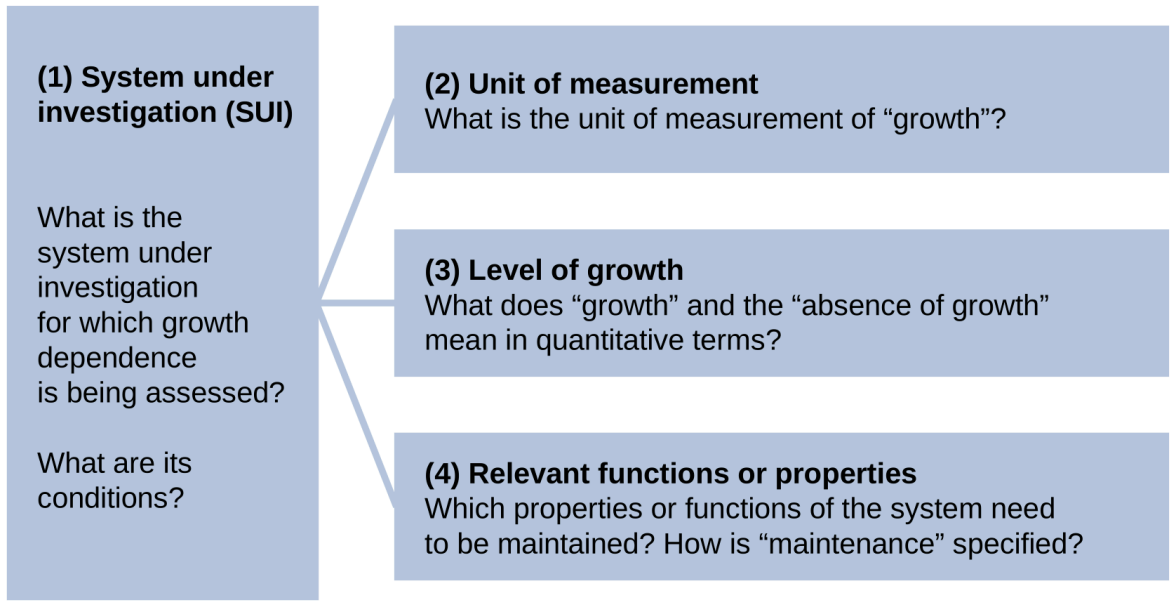

**Figure 1**: Schematic representation of the framework. The aspects (2), (3) and (4) can be understood as defining aspects of growth dependence as characteristic of the system under investigation (1).

## 3.1 System under investigation

First, what is the system being studied? The first step is to delineate the system under investigation whose growth dependence is examined. For instance, such systems could be labor markets, social insurance systems, systems of public finance, stock markets, firms, or national economies. The framework can be applied to different levels and subsystems, from regional economies to the global economy.

Speaking of "systems", rather than "institutions" or "areas of society", makes the concept applicable to a broader range of social and economic structures. According to Meadows (2008, p. 188), a *system* is:

> A set of elements or parts that is coherently organized and interconnected in a pattern or structure that produces a characteristic set of behaviors, often classified as its "function" or "purpose".

While definitions of systems are not uniform (Elder-Vass 2007), Meadows' definition usefully underlines structural features of elements and interconnections, which one can relate to real world elements and interconnections, such as firms and workers being connected through employment relations in labor markets. A particular advantage of the systems perspective is the constitutive distinction between system and system environment, including the forms of exchange between the two (see Luhmann 1995; Mele 2010). This makes it possible, for example, to distinguish the labor market system from social insurance systems without losing sight of their interdependence. Of course, the crucial issue in research design is not so much to exactly define the term "system" as to clarify the system under investigation, including its geographical and temporal coordinates, as well as assumptions about the policies in place that shape growth dependence.

## 3.2 Unit of measurement

Second, what is the measuring unit of "growth" in the term "growth dependence"? While most of the existing literature refers to macroeconomic GDP growth, some analyses also include microeconomic measuring units, such as a firm's revenue, profits or market shares. This aspect is seen as one of five crucial dimensions of growth dependence by Corlet Walker et al. (2024). We focus mainly on real GDP, which is is the inflation-adjusted measure of the market value of all final goods and services produced domestically in a given period[5]. The use of other macroeconomic measuring units, such as GDP per capita, gross national income (GNI), or nominal GDP, may be relevant for specific applications but likely leads to different outcomes of a system's growth dependence assessment, as illustrated in section 4.

Many systems such as firms or public organizations may allow for the distinction between an "inner" measuring unit and an "outer" measuring unit. In the case of social systems, the "inner" measuring unit would be revenues from individual contributions or subsidies, or revenues, market share or profits in case of firms. The "outer" unit then measures the economic activity of a larger system or system environment in which the system under investigation is embedded, such as national GDP in case of national social insurance systems, or global GDP for interbank markets which are embedded in the global economy. Pension systems, too, can be understood to be embedded in the global economy via the link of investment in international financial markets (Wiman 2024). Moreover, the global perspective is relevant for questions such as the impact of lower growth in high-income countries on other parts of the world (see Tan & Tang 2013; Mayer 2017). Figure 2 illustrates this idea of embeddedness.

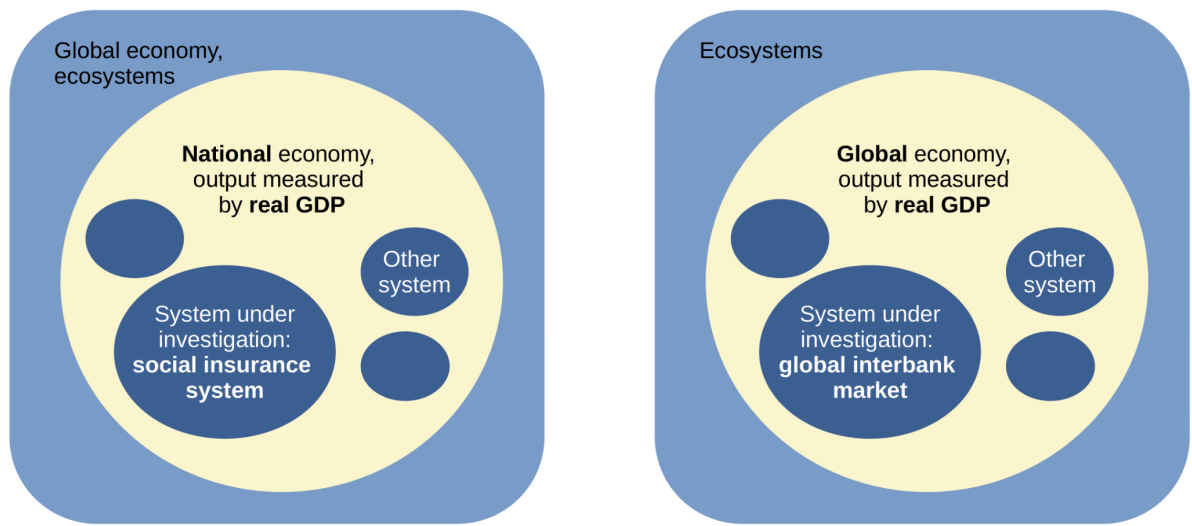

**Figure 2**: Schematic representations of two examples for a system under investigation embedded in a larger system: The social insurance system within the national economy (left panel), and the global interbank market within the global economy (right panel). In both panels, the aspect of the measuring unit refers to real GDP of the larger, "outer" system, such as the national or global economy, embedded in the global economy and/or global ecosystems, respectively.

In sum, the main specification of the measuring unit is real GDP. In some instances, other macro- or micro-level measures may be specified, but the distinction should be made clear.

---

[5] See. e.g. Landefeld (2008) or Campante et al. (2021). Note that the exact definition of GDP is subject to changes over time (see e.g. Christophers 2011 or Semieniuk 2024).

## 3.3 Level of growth

Third, what does "growth" in "growth dependence" mean in quantitative terms? What does the "absence of growth" mean? The term "growth" gives rise to several plausible interpretations with different meanings and different degrees of generality. In our general definition, "growth" denotes a positive growth rate greater than zero over a specified time period. The term "growth rate" captures the relative change of the measuring unit over time, such as an annual percentage rise. For example, a labor market is growth-dependent if a positive growth rate – say 1%, 0.5% or 0.01% – is required in the long run to maintain its relevant property – say, stable employment rates. Consider two hypothetical countries. In the first, employment rates decrease significantly if economic growth is below 1% while in the other the same happens only if economic growth falls below 0.5% per year. According to our general definition, the labor markets in both countries are growth-dependent. Now consider a third country with stable employment rates for any positive and for zero economic growth, and rising unemployment rates only in an economic contraction. This third country is then not growth dependent according to the general definition, since a stable employment rate is maintained in a zero economic growth, and the labor market thus does not require positive growth rates to maintain its relevant property.

Furthermore, two types of effects caused by lower growth can be differentiated: linear vs. non-linear effects. Linear effects of lower growth manifest themselves in the gradual deterioration of relevant functions or properties while non-linear effects unfold through larger and more abrupt diminishments of the system's functions or properties (Corlet Walker et al. 2024, p. 3). While this distinction is relevant for categorizing observed growth dependencies, an operational definition may encompass both types by saying that a system is growth dependent if its functions or properties significantly deteriorate with lower growth, which can happen either gradually or abruptly.

A final possibility for specifying "growth" is to focus on the resilience of a system's relevant functions to the effects of a transition from higher to lower growth or from positive to negative growth (see e.g. O'Neill 2012). Such a transition may pose numerous challenges, for example considering wellbeing (Büchs & Koch 2019), livelihoods (Vogel et al. 2024) or financial stability (Tokic 2012). Such challenges are likely to be more severe than challenges in a long-term zero-growth scenario. The distinction between long-run zero-growth scenarios and transition periods becomes relevant when deciding whether a system requires a sustained transformation to adapt to zero growth, or whether temporary measures for a transition period are sufficient. Restricting the use of the concept "growth dependence" for the long term view helps to make this distinction. Thus, we propose to not use the concept to describe the lack of a system's functionality during transition periods from higher to lower growth, but to refer to the requirement of positive long-run growth rates for the maintenance of relevant functions or properties.

## 3.4 Relevant functions or properties

Fourth, the maintenance of which functions or which properties of the system does one refer to? And how is their non-maintenance specified? Functions of socio-economic systems are for example the provision of income or social inclusion. For instance, in the labor markets, a stable employment rate is an example for a relevant property. The non-maintenance of employment can be then specified as a statistically significant annual rise in the unemployment rate.

From the perspective of systems theory, self-perpetuation is one of the fundamental functions of most systems. If a system's self-perpetuation is dependent on growth, it exhibits an existential form of growth dependence. For instance, one might find that a capitalist firm eventually ceases to exist if it generates zero profits for a while. More often, insufficient growth will not necessarily cause the non-existence of systems. For example, rising unemployment because of low growth does not induce the disappearance of the labor market.

There is no universal way to determine which properties or functions should be considered relevant. As described above, relevance is a normative issue. In line with much of the existing research on growth dependence, we suggest a conscious and transparent choice of a normative framework that focuses, for example, on human well-being. For several different determinations of well-being in normative frameworks, ranging for example from pleasure and desire-fulfillment to needs and capabilities, see e.g. Lamb & Steinberger 2017; Robeyns 2017, p. 118-126.

Because systems are interconnected, low growth in one system might also have repercussions for other systems. For instance, rising unemployment might imply lower revenues for pension systems, which may then impact the wellbeing of people receiving pensions. For analytical clarity, the outcome of lower pensions should not determine whether one considers the labor market itself to be growth-dependent, unless the payment of social benefits is explicitly mentioned as a function of labor markets. Thus, non-maintained functions or properties of other socio-economic systems should not define growth dependence of the system under investigation.

To summarize, if the functions or properties relevant within the chosen normative framework are maintained in the absence of growth as specified by aspect (3), the system is not growth-dependent according to our definition. Conversely, if at least one of these predefined functions or properties drops below a predefined level, the system in question is growth-dependent.

## 3.5 Summary

The four aspects described above form the components of rigorous definitions of growth dependence: (1) *A system under investigation is called growth-dependent if (3) insufficient or lower growth of (2) the specified measuring unit results in (4) at least one relevant function or property to not be maintained.* Apart from a broader understanding in the third aspect, this definition is conceptually the same as the general definition at the start of Section 3.

Conversely, if a system is growth-independent, all of its relevant functions are maintained in the absence of growth as measured. However, even if a system is growth-independent with regard to one growth measuring unit (such as GDP) it might still be growth-dependent with regard to another growth measuring unit (such as sector-level production input). Thus, the identification of a system as growth-independent requires the specification of the complete set of all relevant functions and is limited to the specified growth measuring unit. In Table 1, we present an overview of definition specifications.

**Table 1**: Overview table of the framework, including possible specifications as well as suggestions of what not to include in the umbrella term of growth dependence.

| Aspects defining growth dependence of the system under investigation: | Unit of measurement | Level of growth + no-growth | Relevant functions and properties + non-maintenance |
|---|---|---|---|
| **Specifications:** | The unit of measurement in the concept of growth dependence is mainly specified as … <br><br> … real GDP; <br><br> Other specifications are <br><br> … per capita GDP; <br><br> … nominal GDP; <br><br> … firm's revenue, market share, profits. | The system under investigation is growth dependent, if <br><br> … it requires a specific positive growth rate to maintain its relevant functions or properties; <br><br> … lower levels in the growth rate imply a worse outcome for the relevant functions or properties compared to higher levels of growth. | The non-maintenance of relevant functions or properties can mean … <br><br> … existential or non-existential consequences for the system under investigation; <br><br> … a drop below a certain level of tolerance; <br><br> … a declining dynamic over time. |
| **Excluded specifications:** | | Phenomena in which undesired consequences only occur in a transition from higher to lower growth, but not in zero-growth or when comparing high and low (long term) growth scenarios should not be called growth dependence. | Non-maintained functions or properties of other socio-economic systems should not define growth dependence of the system under investigation. |

# 4 Definitions of growth dependence for employment, social insurance and public finance

The concept of "growth dependence" is prominently adopted in research on employment, social insurance systems and public finance. For each of these systems under investigation, different specifications of the measuring unit of growth, the level of growth or the relevant functions or properties of the system can lead to different assessments of its growth dependence. In this section, we illustrate our framework by applying it to the three cases to show how clearer and more consistent research on growth dependence can be produced. We discuss how different definitions lead to different conclusions about a system's growth dependence, and sometimes also to different evaluations of policies aimed at reducing growth dependence or at achieving growth independence: For the case of employment, we discuss two different specifications of the system's functions; for the case of social insurance systems, we discuss two different specifications of the measuring unit of growth; and for the case of public finance, we discuss two definitions that use different specifications of the system's functions and the measuring unit.

## 4.1 Employment

Increasing unemployment is a major concern regarding non-growing economies (see Antal 2014). We discuss two definitions, differing in how the relevant functions or properties – the fourth aspect of our framework – are specified. The first definition holds that a labor market is growth-dependent if it requires positive economic growth to maintain stable employment rates. According to the second definition, a labor market is growth-dependent if it requires positive economic growth to maintain *all* its relevant societal functions.

In both definitions, the *system under investigation* is the formal labor market. In this system, workers and employers are connected via the wage relation. Informal or unpaid labor is not included in the system of investigation. The *measuring unit* of growth is real GDP. The specification of the sufficient or insufficient *level of growth* takes a long-term perspective, not considering temporary fluctuations, such as temporary recessions, but long-run growth rates. Finally, the potential *relevant functions* of labor markets are manifold. Three stand out in modern capitalist societies. First, most peoples' livelihoods depend directly or indirectly on wage incomes. Second, access to social security benefits is often tied to employment. Third, employment is a major source of inclusion, recognition and personal identity (Frayne 2015).

The employment rate is a property of the system that can serve as an approximation of its ability to fulfill the above-stated functions. The first definition therefore revolves around the employment rate as a proxy for all the relevant functions. The second definition, by contrast, refers explicitly to all relevant functions. In this latter definition, a labor market is growth-dependent if any one of these functions is no longer maintained because of the absence of long-run positive real GDP growth, independently of whether and how the unemployment rate evolves. Using these definitions consciously can improve the assessment of the growth dependence of labor markets.

Numerous studies implicitly apply the first definition. These studies examine the empirical relationship between real GDP growth and the unemployment rate. Most have found a negative relationship, known as Okun's law (see e.g. Ball et al. 2017), with only scattered counterexamples (see e.g. Conteh 2021). More specifically, econometric analyses show that unemployment tends to rise in the case of zero economic growth rates (Sögner & Stiassny 2002; Ball et al. 2017; see Antal 2014 for an overview). The explanation is that if labor productivity increases, the same economic output can be achieved with less labor input. As a result, output growth is required to prevent employment from declining (Antal 2014).

Policy measures for reducing the growth dependence of labor markets are abundant in this strand of research. They often focus on redirecting technological change from labor productivity growth toward improving energy and resource productivity (Lange 2018; Petschow et al. 2020), thereby intensifying economic activity in sectors with low productivity (growth) (Jackson & Victor 2011), or shifting the cost of labor to energy and resources (Köppl & Schratzenstaller 2021). Other proposed policy measures include working time reductions (Kallis et al. 2013), a job guarantee (Alcott 2013), universal basic income (Weeks 2011) and universal basic services (Gough 2019). While some of these policies also aim to maintain stable employment rates (e.g., working time reductions, job guarantee), these policies tend to have broader goals, which are more in line with the second, broader conceptualization of labor markets' functions: These policies try to decouple employment from some of its social functions, such as the provision of income (e.g., universal basic income), livelihoods (e.g., universal basic services) or social status (job guarantee). Thus, the specification of the labor market's functions in the definition of growth dependence matters for the design and evaluation of policy measures aimed at reducing growth dependence.

In sum, a labor market can be said to be growth-dependent if it requires positive economic growth rates to maintain stable employment rates or, alternatively, all of its social functions. The suitability of policies to alleviate the respective dependencies depends on the chosen system functions.

## 4.2 Social insurance

A second system that many scholars consider to be dependent on economic growth is the social insurance system (Bohnenberger 2023). The potential collapse of social insurance systems in the absence of growth is a major concern in academic and broader public discourse. As before, we discuss the specifications of the framework along two definition examples.

In the first definition, a social insurance system is growth-dependent if it requires positive growth in individual contributions or other forms of income to provide the expected level and scope of welfare provisioning. In a second definition, a social insurance system is growth-dependent if it requires positive real GDP growth to provide the expected level and scope of welfare provisioning.

As *the system under investigation*, the welfare system entails various modes of provisioning, including public and private insurances against unemployment, disability and sickness, as well as pension funds and long-term care systems (Garland 2014; Rothgang 2010). These variations of welfare systems are based on mandatory and voluntary financial contributions, which in turn imply an individual's entitlement to benefits (Clegg 2018; Garland 2014; Obinger 2021). Although they are primarily financed through individuals' contributions, they are sometimes "subsidized" by the government (Morel & Palme 2018), which creates a link between the social insurance system and the system of public finance (see next section).

The growth dependence of social insurance systems can be assessed with reference to an "inner" or an "outer" *measuring unit* of growth: the inner measuring unit is the level of revenues from individual contributions or subsidies, and the outer measuring unit is the real output of the economic system in which the respective social insurance system is embedded, measured as real GDP. In both definitions, the required *level of growth* is again understood in terms of general positive long-run growth rates. Lastly, the system's *relevant functions or properties* depend on expectations that both the contributors and beneficiaries have towards social insurance systems. These expectations are highly context-dependent and heterogeneous over time and space (Roosma et al. 2013; Wulfgramm & Starke 2017). The level of welfare provisioning through the social insurance system is determined by the scope of risks covered and by the recipient groups while the total expenditures are determined via the monetary cost of providing these benefits. Through these total costs, the level of welfare provisioning that is expected from the system – its function – is connected to the level of growth upon which the system is dependent.

Once more, any assessment of the growth dependence of social insurance systems depends on the chosen specifications of our framework's four aspects. Notably, Wiman (2024) shows how growth dependence differs across various pension plan types, that is, different types of the *system under investigation*. The author discusses that funded pension schemes, which rely on achieving sufficient rates of return on investment of funds by member contributors, are more likely to be growth dependent than unfunded schemes that finance pensions via redistribution of incomes from the current workforce. Furthermore, Höpflinger (2010) and Petschow et al. (2020, p. 96-100) argue that, in the context of Germany and other European countries, the pension system is growth-dependent with respect to what we term the inner measuring unit of growth, as demographic change (decreasing birth rates and increasing life expectancy) leads to increasing costs of the level of service provision expected from the pension system and thus requires growing contributions. Similar mechanisms apply to health insurance, with further complications resulting from cost-increasing technological developments and a potential Baumol's cost disease (Studer 2010; Petschow et al. 2020, p. 99-100). Here, whereas inner measures of growth do not consider potential distributional conflicts about the sources of contributions and subsidies for pensions and health care, outer measures include these distributional conflicts, for example between spending on unemployment benefits and investments. This difference could lead to diverging assessments of dependence on growth.

In the case of social insurance systems, different specifications of the *functions or properties* of the system significantly influence the assessment of whether a growth dependence exists or not. As in the other cases, insurance systems and their growth dependence are politically determined in the sense that the system's level and scope of provisioning, as well as its revenue sources, are subject to political conflict.

## 4.3 Public finance

The notion that public finance is dependent on GDP growth is widespread. Many commentators assume that a certain level of GDP is required to "finance" a certain level of government expenditure, and that "debt sustainability" requires GDP growth rates at least on par with the interest rate on public debt. This concern is often raised by skeptics of post-growth (e.g., Pollin 2019; Pasche 2018), but shared also by some post-growth advocates (Seidl & Zahrnt 2010, p. 16; Salama 2023). However, no systematic scientific analysis known to us has substantiated these assumptions of a blanket growth dependence of public finance. More rigorous analysis starts with more precise definitions. Different specifications of the four aspects in our framework can lead to very different notions of what growth dependence means in the case of public finance.

To illustrate, we present two different definitions. In a narrow *legal* definition, public finance is growth-dependent if a government is unable to observe legal debt rules while providing adequate public services in the absence of positive nominal GDP growth rates. In a broader *political-economic* conception, public finance is growth-dependent if a government is unable to provide an acceptable level and scope of public services, ensure low inflation, and choose fiscal policies with politically accepted distributional outcomes in the absence of positive real GDP growth.

Public finance, the *system under investigation*, can be defined as the practices and policies that central and local governments use to govern their expenditure, their revenue, and their debt. The *measuring unit* can be specified as nominal or real GDP, depending on whether inflation is considered as relevant to the functions of public finance. Accordingly, the *level of growth* refers to a positive growth rate of nominal or real GDP. Finally, the *relevant functions* of public finance are context-dependent and politically contested, but typically include some tradeoff between macroeconomic stability, legal debt rules, distributional outcomes and the provision of public services.

The *functions* of public finance in the first definition are first, to provide adequate public goods while observing legal constraints the ratio of government deficits or debts to nominal GDP (e.g., the German and Swiss constitutional debt brakes or the Maastricht criteria). In a scenario where the interest rate on public debt is higher than the nominal GDP growth rate, the debt-to-GDP ratio grows even if the primary government budget is balanced (see Pasche 2018). Thus, nominal GDP growth at least on par with the interest rate is necessary to ensure that both functions of public finance – providing public goods and observing legal debt limits – can be fulfilled at the same time. Under the first definition and some further assumptions (see Pasche 2018), any public finance system with legal debt rules is growth-dependent in a scenario with positive nominal interest rates.

The second definition refers to broader political-economic *functions* of public finance, encompassing again the provision of adequate public goods, but also ensuring low and stable rates of inflation, and adopting fiscal policies whose distributional outcomes are politically acceptable. In a scenario where exogenous developments (for instance, demographic change as explained in the previous section) require rising government expenditure to provide the same level of public goods as before, there must be either real GDP growth, inflation or contractionary fiscal policies. The reason is that higher government expenditure implies an increase in aggregate demand, which normally drives either an increase in aggregate supply – that is, real GDP growth – or, if supply is constrained, inflation (see Olk et al. 2023). Fighting inflation may require politically unacceptable fiscal (and monetary) policies. In this scenario, there is a conflict between the different functions of public finance – adequate public services, low inflation, and acceptable fiscal policies – that can be resolved only through real GDP growth.

# 5. Discussion

The application of the presented growth dependence framework in the three examples highlights the key insight for further research and policy design: It depends. Whether social-economic systems, such as employment, social insurance, or public finance, are growth-dependent depends not only on the specific properties of these systems, but also on the choice of their operationalization. Research on growth (in)dependence and the public debates about some of the core issues surrounding the sustainability agenda could benefit immensely from adopting a coherent framework for the operationalization of growth dependence.

In many analyses, labor markets are linked to economic growth, yet the exact causal relations are contested as we discuss by highlighting two divergent specifications of the system's functions: either maintaining stable employment rates in the absence of growth (as is often discussed with reference to Okun's law) or maintaining specific social functions such as providing for peoples' livelihoods, social security, or social recognition, independently of whether and how this is reflected in the unemployment rate. Similarly, growth dependence also depends on specifying measuring units, as we discussed with reference to social insurance systems. Social insurance systems are growth-dependent if, to provide the expected level and scope of welfare provisioning, they either require positive growth in individual contributions or subsidies, or if they require positive real GDP growth. Only analyses that are based on the latter definition can account for potential distributional conflicts between funding sources. Finally, for the case of public finance, two important definitions of growth dependencies are differentiated not only in their specifications of the measuring unit, but also the system's functions. In a narrow legal definition, public finance is growth-dependent if without positive nominal GDP growth, a government is unable to observe legal debt rules while providing adequate public services. While this definition can inform current debates around austerity and public finance in the context of low growth, a broader political-economic conception of growth dependencies is key for understanding the challenges of post-growth public finance: In this understanding, public finance is growth-dependent if because of a lack of positive real GDP growth, a government is unable not only to provide an acceptable level and scope of public services, but also to ensure low inflation, and to choose justice-oriented fiscal policies.

Beyond the three examples, our framework for analyzing growth dependencies could be applied to several other areas that are still underexplored, in particular stability of financial markets, distributional dynamics or questions regarding the functioning of decentralized markets in a zero-growth economy in general. Respective questions are: Does the stability of financial markets depend on real or expected economic growth rates (see e.g. Tokic 2012)? Under which conditions does inequality rise in a zero-growth economy, and specifically for which types of inequalities (see Jackson & Victor 2016; Janischewski 2022)? Do decentralized markets function for various goods and services in a zero-growth economy, and under what conditions? More generally, our framework could help researchers to explore systematically which provisioning systems are growth-dependent, defined as a "set of related elements that work together in the transformation of resources to satisfy a foreseen human need" (Fanning et al. 2020, p. 1). Moreover, the presented framework allows for the analysis of cultural, psychological or ideological causes of growth dependencies.

Last but not least, while our illustrations have focused on three separate systems, growth dependencies do not exist in isolation. In fact, socio-economic systems are interrelated and thus should not be addressed separately. For example, increasing unemployment (the labor market system) as a result of absent GDP growth means that tax revenues decrease (the public finance system), which in turn increases the need for public spending in the form of unemployment benefits (the social insurance system) (Hirvilammi 2020). Furthermore, if public spending has to provide for growing needs in social benefits not covered by the social insurance systems in a zero-growth economy, the respective cause of growth dependence in social insurance systems is shifted to public finance. Similarly, policies in the context of employment such as a green job guarantee on the one hand tackle the issue of rises in unemployment but on the other hand increase the requirements on public spending, and can thus enhance potential growth dependencies of the latter system. In conclusion, while the identification of growth dependent areas of society requires both, distinct analyses of the various growth dependencies and analyzing their interconnections, the design of measures for the alleviation of growth dependence needs a holistic perspective that takes into account inter-system feedbacks and dependencies.

# 6. Conclusion

Growth dependencies in various socio-economic systems of modern societies are key barriers to achieving human flourishing without further destroying planetary ecosystems. However, dismantling such barriers requires tools for their analyses and schemes for their unambiguous identification. To catalyze in-depth and (comparative) research of growth-dependent systems and policies for increasing growth independence, we have introduced a general definition and a novel framework encompassing four simple elements: (1) the specification of the system under investigation, (2) the unit of measurement, (3) the level of "growth" that we refer to in the term "growth dependence", and (4) the system's relevant functions or properties whose maintenance is essential for human well-being. Our illustrative discussion yields an important message about the possibility of appraising growth dependence: It depends on careful operationalization.

We have demonstrated the framework's validity through a threefold illustrative application to employment, social insurance and public finance. This way, we provide a scientific basis for further explorations of growth dependence across various socio-economic systems, supporting the interdisciplinary comparability of research results. The framework forms a basis not only for the analysis and comparison of existing socio-economic systems across geographical and temporal dimensions, but crucially also for the assessment of the capacity of policy proposals to mitigate growth dependencies. Answering questions about the extent and features of the growth dependence of socio-economic systems is crucial to designing policy measures that can help to overcome these dependencies – and thus a prerequisite for reaching and then safeguarding socio-ecological sustainability.

## Acknowledgements

Earlier drafts were presented at the ICTA-UAB Growth vs Climate Conference 2024 in Barcelona, March 13 - 15 2024; and the joint Conference of the European Society for Ecological Economics and the International Degrowth Conference 2024 in Pontevedra, June 18 - 21, 2024. We wish to thank the audiences at these events for engaging with the ideas developed in this paper. We furthermore thank Torsten Heinrich, David Hofmann, Jenny Lay-Kumar, Ulrich Petschow, Benedikt Schmid and Irmi Seidl for their helpful comments and suggestions. Furthermore, we are grateful to Daniel Eichhorn for facilitating the collaboration process in the first months.

## CRediT authorship contribution statement

**Anja Janischewski**: Conceptualization, Writing - Original Draft, Writing - Review & Editing, Project administration, Visualization; **Katharina Bohnenberger**: Conceptualization, Writing - Original Draft, Writing - Review & Editing; **Matthias Kranke**: Writing - Original Draft, Writing - Review & Editing; **Tobias Vogel**: Conceptualization, Writing - Original Draft, Writing - Review & Editing; **Riwan Driouich:** Writing - Original Draft, Visualization; **Tobias Froese:** Writing - Original Draft; **Stefanie Gerold**: Writing - Original Draft; **Raphael Kaufmann**: Writing - Original Draft; **Lorenz Keyßer**: Writing - Original Draft, **Jannis Niethammer**: Writing - Original Draft, Resources; **Christopher Olk**: Writing - Original Draft, Writing - Review & Editing; **Matthias Schmelzer:** Writing - Original Draft; **Aslı Yürük:** Writing - Original Draft, Visualization; **Steffen Lange**: Conceptualization, Writing - Original Draft, Writing - Review & Editing

**Core group**: Anja Janischewski, Katharina Bohnenberger, Matthias Kranke, Steffen Lange, Tobias Vogel;
**Contributing authors**: Riwan Driouich, Tobias Froese, Stefanie Gerold, Raphael Kaufmann, Lorenz Keyßer, Jannis Niethammer, Christopher Olk, Matthias Schmelzer, Aslı Yürük

# Funding statement

M.K. acknowledges financial support by the Federal Ministry of Education and Research through a Senior Research Fellowship at the Käte Hamburger Kolleg/Centre for Global Cooperation Research, University of Duisburg-Essen, Germany [grant number 01UK1810]; and by the Eva Mayr-Stihl Foundation through a Young Academy for Sustainability Research Fellowship at the Freiburg Institute for Advanced Studies (FRIAS), University of Freiburg, Germany.

# Appendix A

**Table A.1**: Non-comprehensive list of growth (in)dependence definitions and descriptions.

| Defining statement on growth (in)dependence | | Reference |
|---|---|---|
| **Growth dependence** | "By 'dependent' we mean that welfare systems require the continuation of economic growth in order to avoid significant negative social and economic consequences." | Bohnenberger & Fritz (2020, p. 4) |
| | "… modern economies have become 'growth dependent' in the sense that certain core aspects of human wellbeing become compromised when, for whatever reason, growth in the GDP is either hard to come by or is undesirable." | Corlet Walker & Jackson (2021, p. 2) |
| | "Growth dependence can be broadly thought of as those conditions that require the continuation of economic growth in order to avoid significant psychological, social, and economic harms (e.g. mass unemployment, poor health outcomes, etc.). The specific form that growth dependencies take varies from system to system." | Corlet Walker et al. (2021, p. 5) |
| | "To aid this task we begin with the following working definition of growth dependency: *Those conditions that require the continuation of growth in order to avoid significant physical, psychological, social, and/or economic harms*." | Corlet Walker et al. (2024, p. 2, original emphasis) |
| | "In this discussion paper we define as growth-dependent those areas of society, structures, institutions, etc.<br>► that fulfil a socially desirable function, or that contribute to a widely socially accepted objective and<br>► whose functional capacity or contribution under the present framework conditions depends on the economy growing continually." | Petschow et al. (2020, p. 91) |
| | "Economic structures, social institutions and even the subjects of capitalist societies are fundamentally oriented towards growth and can therefore only be 'dynamically stabilised'…. In other words, institutions, infrastructures and subjectivities can only be maintained in their prevailing constitution by continued growth." | Schmid (2021, p. 65) |
| | "One reason is that several societal areas are growth-dependent. These areas fulfil a socially desirable function and contribute to an important societal goal. But under current conditions, their functionality and contribution to society depend on continuous economic growth ….." | Wagner & Lange (2021, p. 44) |
| | "… *adverse economic effects from the end of growth or the abandonment of a growth assumption*, indicating growth dependence." | Wiman (2024, p.2, original emphasis) |

| **Growth independence** | "To that end, an entirely different mode of production and consumption is deemed necessary, which requires a fundamental transformation of the growth-based societal and economic institutions, mindsets, and infrastructures .... Economic growth, ultimately, would no longer be the structurally fixed and culturally engrained necessary condition for economic stability and human wellbeing ...." | Gebauer (2018, p. 232) |
|---|---|---|
| | "An important idea here among post-growth theorists is that of 'growth independence' .... It is recognised that many social benefits (notably employment, pensions and public services) are currently dependent on growth for their achievement and improvement. So a key policy goal is to make such benefits independent of growth." | Likaj et al. (2022, p. 27) |
| | "Growth independence can be defined as the ability of a society, including its economy and its institutions, to continue to fulfil its functions, but not to be existentially dependent on economic growth ...." | Mayer et al. (2021, p. 219, original emphasis) |
| | "A degrowth society is a society that, through a democratic process, transforms its institutions and infrastructures so that they are *not dependent on growth and continuous expansion* for their functioning." | Schmelzer et al. (2022, p. 206, original emphasis) |
| | "Growth independence is not understood as the opposite of growth, namely shrinking. We rather adopt the meaning established in the post-growth literature ...: the ability of a society including its economy and its institutions to continue to fulfil its functions but no longer to be existentially dependent on economic growth .... Basic social and economic functions include safeguarding livelihoods, participation in society for all, basic infrastructure and healthcare." | Tschumi et al. (2021, p. 119) |
| | "Environmental policies possibly leading to a decline in economic growth threaten the viability of these areas. Shaping the latter in a way that they can fulfil their socially desirable function even if the economy is not growing would release environmental policies from any reservations regarding limiting growth. In other words, establishing growth-independent areas is necessary if Germany, and other developed countries are to be steered onto a path towards staying within planetary boundaries." | Wagner & Lange (2021, p. 44) |